\documentclass{article}
\usepackage{spconf,amsmath,graphicx,amssymb,subcaption,xcolor}
\usepackage{booktabs}
\usepackage[font=small,belowskip=1pt,aboveskip=2pt]{caption}

\makeatletter
\renewcommand\section{\@startsection {section}{1}{\z@}%
                                	{-3ex \@plus -0.1ex \@minus -.2ex}%
                                	{1.3ex \@plus.12ex}%
                                	{\normalfont\Large\bfseries}}
\renewcommand\subsection{\@startsection{subsection}{2}{\z@}%
                                	{-2ex\@plus -0.1ex \@minus -.2ex}%
                                	{1ex \@plus 0.01ex}%
                                	{\normalfont\large\bfseries}}
\renewcommand\paragraph{\@startsection{paragraph}{5}{\z@}%
                                       {0.5ex \@plus0.2ex \@minus1.01ex}%
                                       {-0.25em}%
                                       {\normalfont\normalsize\bfseries\slshape}}
\renewcommand\subparagraph{\@startsection{subparagraph}{6}{\parindent}%
                                          {0.25ex \@plus0.2ex \@minus .2ex}%
                                          {-0.2em}%
                                          {\normalfont\normalsize}}
\makeatother

\newcommand{\etc}{\textit{etc}}
\newcommand{\ie}{\textit{i.e., }}
\newcommand{\eg}{\textit{e.g., }}

\makeatletter
\newcommand*\titleheader[1]{\gdef\@titleheader{#1}}
\AtBeginDocument{%
	\let\st@red@title\@title
	\def\@title{%
		\bgroup\normalfont\small\@titleheader\par\egroup
		\vskip1.5em\st@red@title}
}
\makeatother

\title{Characterizing dynamically varying acoustic scenes from egocentric audio recordings in workplace setting}
\titleheader{\textit{This is a preprint. The paper is submitted to IEEE International Conference on Acoustics, Speech and Signal Processing (ICASSP), 2020.}\\\rule{\linewidth}{1pt}}
\name{Arindam Jati \quad Amrutha Nadarajan \quad Karel Mundnich \quad Shrikanth Narayanan}
\address{Signal Analysis and Interpretation Lab, University of Southern California, Los Angeles, CA 90089}
\begin{document}
\ninept

\setlength{\abovedisplayskip}{1.5pt}
\setlength{\belowdisplayskip}{1.5pt}

\maketitle
\begin{abstract}
\vspace{-0.5mm}
Devices capable of detecting and categorizing acoustic scenes have numerous applications such as providing context-aware user experiences.
In this paper, we address the task of characterizing acoustic scenes in a workplace setting from audio recordings collected with wearable microphones.
The acoustic scenes, tracked with Bluetooth transceivers, vary dynamically with time from the egocentric perspective of a mobile user.
Our dataset contains \textit{experience sampled} long audio recordings collected from clinical providers in a hospital, who wore the audio badges during multiple work shifts.
To handle the long egocentric recordings, we propose a Time Delay Neural Network~(TDNN)-based segment-level modeling.
The experiments show that TDNN outperforms other models in the acoustic scene classification task.
We investigate the effect of primary speaker's speech in determining acoustic scenes from audio badges, and provide a comparison between performance of different models.
Moreover, we explore the relationship between the sequence of acoustic scenes experienced by the users and the nature of their jobs, and find that the scene sequence predicted by our model tend to possess similar relationship.
The initial promising results reveal numerous research directions for acoustic scene classification via wearable devices as well as egocentric analysis of dynamic acoustic scenes encountered by the users.
\end{abstract}
\begin{keywords}
Acoustic scene classification, audio event detection, dynamic audio environment, time delay neural network
\end{keywords}
\section{Introduction}
\label{sec:intro}
Audio modality often provides information streams that are important in building multi-modal systems such as for enabling context-aware and personalized user experiences.
There is a fast-growing research interest in building systems capable of understanding the ambient acoustic environment: this includes both its dynamically evolving nature in a given location, and across locations such as from the view point of a mobile user.
An ``acoustic scene''~\cite{stowell2015detection} refers to an audio environment characterized by the ``sound events''~\cite{gemmeke2017audio} that occur in it. 
Machines with the ability of identifying the acoustic scene in a given audio recording have numerous applications in robot navigation~\cite{stowell2015detection,chu2006scene}, context-aware devices and associated user notifications, advanced gaming systems, accessibility systems, self-driving vehicles, and surveillance~\cite{atrey2006audio}.
The machine learning task of identifying the acoustic environment in an audio recording is generally known as Acoustic Scene Classification~(ASC)~\cite{stowell2015detection}.
Over the past few years, significant progress has been made in this domain of research in the form of new datasets (\eg the DCASE challenges~\cite{mesaros2018multi}) and novel algorithms~\cite{mesaros2018multi,mesaros2018detection,barchiesi2015acoustic,mesaros2016tut,richard2013overview}.

Acoustic scenes can vary in granularity of their semantic descriptors, and they can be heterogeneous in nature~\cite{eronen2005audio,chu2009environmental,8682341} -- location (indoor, outdoor, \etc.), sources (alarms, chatter, door slams, \etc.) and so on.
For example, the acoustic environment of a workplace such as a \textit{hospital} can have multiple sub-categories (acoustic locales) of interest such as nurse stations, medication rooms, patient rooms, labs, lounges, \etc., each with distinct acoustic ambiences.
Moreover, from the perspective of the employees in a workplace, the dynamically varying acoustic environments might look different for different job-types. 
A nurse in a hospital setting might experience most of the above acoustic scenes in a certain work shift as compared to a lab technician for instance. 
Identifying such dynamically changing acoustic scenes from an egocentric (centered around a certain person) view of the user, and characterizing their temporal patterns can potentially provide insights about the relationship between the acoustic scenes experienced by the employees and nature of their jobs. 
For example, stress patterns related to acoustic environments could be mapped.

In this paper, we address the task of identifying and characterizing dynamically varying acoustic scenes in a workplace setting from egocentric audio recordings obtained through audio recorders worn as badges by individuals~\cite{feng2018tiles}.
There are three fundamental differences between the task at hand and standard ASC tasks~\cite{stowell2015detection,mesaros2018detection}.
First, to get an egocentric view, we employ wearable microphones for audio feature collection~\cite{feng2018tiles}. 
The employees wear their audio recorders throughout multiple work shifts.
The audio badges always capture the voice of the participant wearing the microphone; the speech is recorded at a higher intensity than the background scenes because of the sensor design. 
Second, from the standpoint of the participant, the acoustic scenes might vary dynamically over time (including across locations for ambulatory individuals such as nurses).
Third, the acoustic scenes are from fine-grained classes inside the umbrella of a workplace (more specifically, a large critical care hospital) setting.

The main contributions of the paper are summarized below. 
\begin{itemize}
    \itemsep0em
    \item This work formulates the problem of identifying dynamically varying acoustic scenes in a real world workplace setting.
    \item The data collection setup makes this a one of a kind problem as we have an unobtrusive, \textit{experience sampled}~\cite{larson2014experience} measurements of location and audio from an egocentric point of view.
    \item The present ASC task is constrained with the possible overlay of the user's speech; acoustic scenes have to be identified even when the user is talking.
    \item The effect of foreground speech (the speaker wearing the audio badge) in predicting acoustics scenes from wearable microphones is analyzed.
    \item A deep learning framework based on a Time Delay Neural Network~(TDNN)~\cite{peddinti2015time} model to learn the segment-level acoustic scenes from audio features is proposed.
    \item The temporal characteristics of the acoustic scenes experienced by the users are analyzed with respect to the nature of their occupations.
\end{itemize}

\section{Dataset}\label{sec:Dataset}
The present work is a part of the ``TILES: Tracking IndividuaL performancE with Sensors'' project, which is a part of the IARPA MOSAIC program\footnote[1]{https://www.iarpa.gov/index.php/research-programs/mosaic}.
The goal of this project is to assess the effect of multiple stressors (stressful events in life~\cite{jati2018towards}) on workplace behaviors, affect, and performance of the employees through the use of off-the-shelf wearable sensors.
We collected multi-modal sensory data (audio, physiology, continuous location, \etc.) from $350$ nurses and other direct clinical providers in a critical care hospital\footnote[2]{USC Keck Hospital, Los Angeles, CA, USA.}.
The data was collected through audio badges~\cite{feng2018tiles} developed in-house, which the participants wore during their work shifts.
Each participant went through the data collection procedure in multiple work shifts, each typically lasted from $8$ to $12$ hours. 
The entire dataset was collected over the period of ten weeks.
The dataset contains multiple days of multi-modal data for every participant, thus contains data with longer temporal context as compared to standard ASC tasks like~\cite{mesaros2018multi}.
This rich context inspired us to deploy segment-level modeling as described in Section~\ref{sec:Methodology}. More details about the dataset can be found in~\cite{booth2019multimodal}.
For this work (in the initial phase), we employ a subset of $86$ participants ($29$ males and $57$ females).

\subsection{Acoustic features}\label{sec:Acoustic features}
In compliance with HIPAA regulations~\cite{anthony2014institutionalizing} and because of the sensitive nature of the study environment (hospital), we were unable to collect raw audio signal.
The audio badge~\cite{feng2018tiles}, equipped with an energy based voice activity detector, collected $125$ low-level descriptor features using the OpenSMILE toolkit~\cite{eyben2010opensmile} at a sampling rate of $16$~kHz.
The features were computed using a moving window of $60$~ms length with $50$~ms overlap~\cite{nadarajan2019speaker}.
The feature set consists of spectral features such as MFCCs, other speech features including pitch and loudness, and voice quality measures like vocal shimmer and jitter. 
We incorporate all features for our current analysis.

\subsection{Acoustic scene location labels}
\label{sec:Acoustic scene labels}
\begin{figure}
    \centering
    \begin{subfigure}[b]{0.48\linewidth}
        \includegraphics[width=\linewidth]{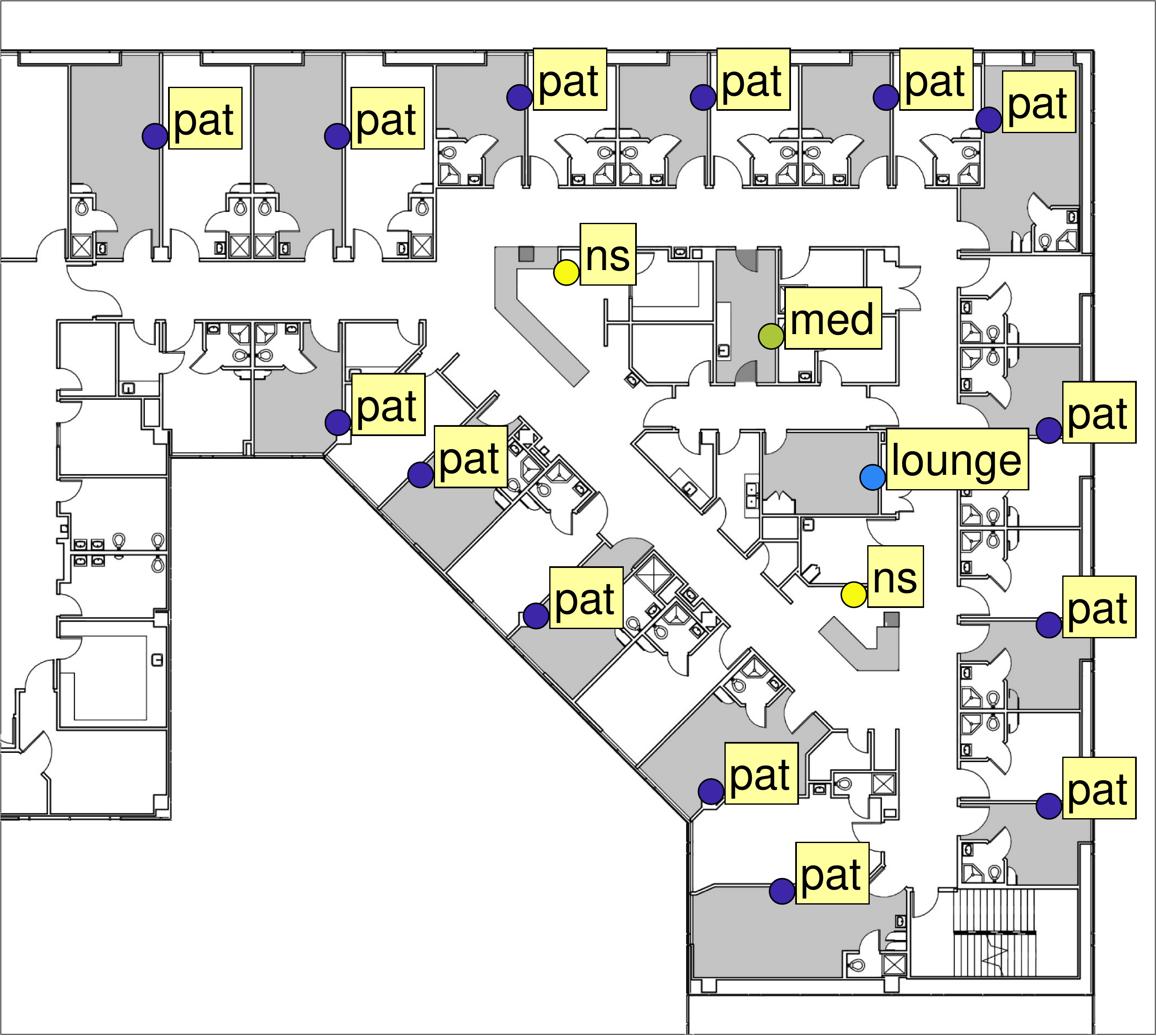}
        \caption{Typical Floor plan.}
        \label{fig:floor}
    \end{subfigure}
    \begin{subfigure}[b]{0.48\linewidth}
        \includegraphics[trim=100 200 100 200,clip,width=\linewidth]{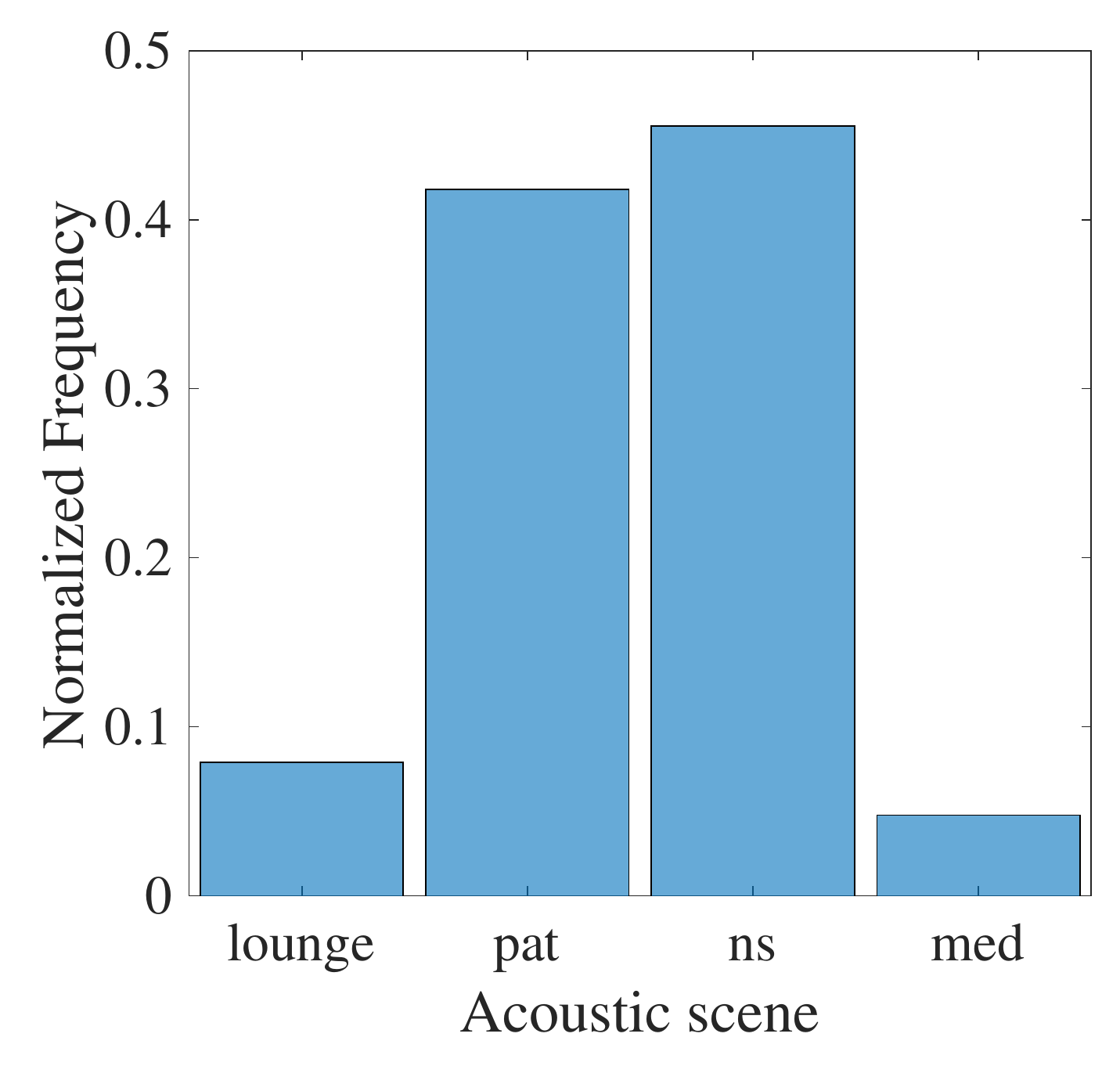}
        \caption{Histogram of scenes.}
        \label{fig:scene_hist}
    \end{subfigure}
    \caption{(a) One of the hospital floor plans showing different acoustic locales. (b) Histogram of acoustic scene samples in our dataset.}
    \label{fig:hist_labels}
\end{figure}

The acoustic scene locales are derived from Bluetooth transceivers installed in different locations in the hospital~\cite{mundnich2019bluetooth} as a part of the TILES study.
The transceivers receive Bluetooth pings sent by the audio badges, and provide Received Signal Strength Indicator~(RSSI) values which are used to track the temporally varying acoustic environment from the participant's perspective.
The temporal resolution of this location data is much coarser ($\sim 1$ minute) than that of the acoustic features, and this needs further alignment step as discussed in Section~\ref{sec:Mining samples from continuous audio}.
At every timestamp, maximum RSSI value is used to determine the fine-grained location of the participant.
We refer the readers to~\cite{mundnich2019bluetooth} for further details about the indoor localization process.
The fine-grained locations are then processed and clustered according to their associated acoustic environments.
In this work, we target four different locations (see Figure~\ref{fig:hist_labels}) in a hospital unit, each having unique characteristics in their acoustic scenes: \textit{nurse stations} (\textbf{`ns'}), \textit{patient rooms} (\textbf{`pat'}), \textit{medication rooms} (\textbf{`med'}), and \textit{lounge} (\textbf{`lounge'}).

\section{Methodology}\label{sec:Methodology}
Because of the having extremely long audio recordings for every participant (see Section~\ref{sec:Dataset}), we address the problem with a segment-level modeling.
The goal is to learn an acoustic scene model that can predict the scene given an input audio (features) segment.
Moreover, we characterize the temporal sequence of predicted acoustic scenes for the subsequent analysis of its relationship with the nature of job..

\subsection{Problem Formulation}\label{sec:Problem Formulation}
Let, $\mathcal{D}=\left\{ \mathcal{X}_i, \mathcal{Y}_i \right\}_{i=1}^N$ be a dataset of $N$ participants.
Here $\mathcal{X}_i$ denotes the temporal sequence of segmented audio features for the $i^\text{th}$ participant:
\begin{equation}
    \mathcal{X}_i = \left[ \mathbf{X}_{i1}, \mathbf{X}_{i2}, \dots, \mathbf{X}_{iT_i} \right].
\end{equation}
Here, $\mathbf{X}_{ij} \in \mathbb{R}^{L_{ij} \times F}$ denotes the $j^{\text{th}}$ segment of the $i^\text{th}$ participant.
It is a 2D matrix containing $L_{ij}$ acoustic feature vectors of dimension $F$. 
For simplicity, in this work we fix the segment length, \ie $L_{ij} = L$.
Therefore, $\mathbf{X}_{ij}$ can be represented as:
\begin{equation}\label{eq:X_ij}
    \mathbf{X}_{ij} = \left[ \mathbf{x}^T_{ij1} | \mathbf{x}^T_{ij2} | \dots | \mathbf{x}^T_{ijL}  \right]^T,
\end{equation}
where $\mathbf{x}_{ijk} \in \mathbb{R}^{F}$ is the $k^{\text{th}}$ feature vector of dimension $F$.
$\mathcal{Y}_i$ denotes the sequence of acoustic scene labels for the $i^\text{th}$ participant:
\begin{equation}
    \mathcal{Y}_i = \left[ y_{i1}, y_{i2}, \dots, y_{iT_i} \right],
\end{equation}
where $y_{ij} \in \{1,2,\dots,C\}$ denotes one of the $C$ acoustic scenes.

\subsection{Acoustic scene modeling}
The acoustic scene predictor is learned on the segment-level acoustic feature streams in a speaker-agnostic setting.
More formally, given $\mathcal{D}$, the task is to learn a nonlinear mapping $f(\cdot)$:
\begin{equation}\label{eq:y_f}
    \hat{y}_{ij} = f\left( \mathbf{X}_{ij}; \Theta \right).
\end{equation}
Here, $f(\cdot)$ is modeled by a Deep Neural Network~(DNN) with parameter set $\Theta$, and $\hat{y}_{ij}$ gives the predicted class label for $\mathbf{X}_{ij}$.
Standard cross-entropy loss is employed as the minimization objective.

\subsection{Time Delay Neural Network~(TDNN)}
Time delay Neural Networks~\cite{waibel1989phoneme} have been found to achieve state-of-the-art performance for speech recognition~\cite{peddinti2015time} and speaker recognition~\cite{snyder2018x} tasks.
TDNNs are conceptually similar to 1D \textit{dilated}~\cite{oord2016wavenet} convolutional neural networks, and thus, they can model long term temporal dependencies with much fewer parameters compared to recurrent neural networks~\cite{peddinti2015time}.
We adopt the architecture of \cite{snyder2018x}, but with minor modifications (details about the parameters are in Section~\ref{sec:Model parameters}).
The TDNN model takes an acoustic feature segment, $\mathbf{X}_{ij}$ (see Equation~\ref{eq:X_ij}), and transforms it to a sequence of embedding vectors through a series of  hierarchical dilated 1D convolution operations:
\begin{equation}
    \left[ \mathbf{z}_{ij1} | \mathbf{z}_{ij2} | \dots | \mathbf{z}_{ijL^{'}} \right] = g\left( \mathbf{X}_{ij} \right ).
\end{equation}
Then, we compute the sample mean of the embedding vectors, and pass them through two more layers of dense transformation (similar to~\cite{snyder2018x}), before reaching the penultimate layer with $C$ outputs having softmax activations.
Similar to Equation~\ref{eq:y_f}, the overall mapping can be expressed as:
\begin{equation}
    \hat{y}_{ij} = f\left( \mathbf{X}_{ij} \right) = h\left( \mathbb{E}_T\left[ g\left( \mathbf{X}_{ij} \right) \right ] \right),
\end{equation}
where $\mathbb{E}_T[\cdot]$ denotes temporal mean function, and $h(\cdot)$ indicates the transformation after the TDNN layers.

\subsection{Characterizing dynamically varying acoustic scenes}\label{sec:Characterizing dynamically varying acoustic scenes}

From the egocentric perspective of a participant, the acoustic scenes may dynamically vary depending on the nature of their job.
The output of a pre-trained model on all the segment-level acoustic features of a participant are temporally ordered to produce the predicted acoustic scene vector:
\begin{equation}
    \mathcal{\hat{Y}}_i = \left[ \hat{y}_{i1}, \hat{y}_{i2}, \dots, \hat{y}_{iT_i} \right].
\end{equation}
We look at how frequently the acoustic scene changes for a particular participant, and if that characteristic is related to the nature their jobs.
Formally, we measure the number of non-zero elements of the following difference signal:
\begin{equation}
    \hat{\delta}_i[t] = \mathcal{\hat{Y}}_i\left[t\right] - \mathcal{\hat{Y}}_i\left[t-1\right].
\end{equation}
Therefore, number of changes in $\hat{\delta}_i$ is given by (normalized with respect to its length):
\begin{equation}\label{eq:nnz}
    \Delta \mathcal{\hat{Y}}_i  = \frac{1}{T_i}\sum_{t=1}^{T_i} \mathbb{I}\left( \hat{\delta}_i[t] \ne 0 \right),
\end{equation}
where $\mathbb{I}(\cdot)$ is the indicator function.
Similarly, we can compute $\Delta \mathcal{Y}_i $ for the true scene sequence, and compare if the information captured by the true sequnce is also captured by the predicted one (Section~\ref{sec:Characterizing predicted sequence of acoustic scenes}).



\section{Experimental Setting}

\subsection{Mining samples from continuous audio}
\label{sec:Mining samples from continuous audio}
Because the difference in the resolutions of audio and location data (Section~\ref{sec:Acoustic scene labels}), we mine $5s$ (\ie $L$ $=$ $500$ frames) audio feature segments from the dataset when there is a location label available.
The sampling method results in a reduction of the amount of audio features to process.
To restrict the model from getting biased toward specific speakers, we normalize every segment of a certain participant by subtracting the mean feature vector computed on that participant's entire data (might be spanning multiple days).
Thus, $500\times 125$ dimensional samples are fed to the acoustic scene models (except the baseline model, see Section~\ref{sec:Model parameters}) for training and inference.

\subsection{Distribution of acoustic scene labels}\label{sec:Acoustic scene label distribution}
Figure~\ref{fig:scene_hist} plots the histogram of all the collected acoustic scene labels in our data.
We can see that the class distribution is skewed.
Most of the samples come from nurse stations and patient rooms.
This makes the problem more challenging since the unbalanced dataset might include some class-specific biases during the training process.
Note that the accuracy of a \textit{majority guess baseline classifier} would be $\sim 46\%$ (\ie percentage of `ns' samples).

\subsection{Effect of foreground speech}\label{sec:Effect of background audio}
To analyze the effect of foreground (FG as abbreviation) speech in predicting the acoustic scenes, we apply a pre-trained foreground detection model~\cite{nadarajan2019speaker} on the audio features.
The foreground detection model was trained in a supervised way (on an out-of-domain corpus containing speech from meetings) to detect the foreground speaker \ie the speaker wearing the close-talk microphone.
This analysis is particularly important here due the usage of wearable devices for collecting the audio features.
We now have two data subsets:
\begin{enumerate}
    \item \textbf{FG active:} Dataset created by mining samples from audio when foreground speaker is supposedly active. It should capture background audio mixed with foreground speech. This has $\sim 64$k samples (train and test).
    \item \textbf{Full:} Dataset generated by sampling from the raw features without applying any FG detection masks. It should capture background audio in presence and absence of the FG speech. This subset contains  $\sim 185$k samples.
\end{enumerate}
Note that the label distribution (Section~\ref{sec:Acoustic scene label distribution}) is almost similar for the two subsets.
We analyze performance of different models separately on these two data subsets.

\subsection{Data splits}\label{sec:Data parameters, splits}
We do $10$-fold cross validation to report test performances.
We ensure that a participant only lies in one out of the $10$ folds, so that the model does not get biased toward certain speakers.
For each test fold, we perform model selection by utilizing a validation set curated from $4$ participants in the train set (remaining $9$ folds).
We choose the model with lowest validation loss for evaluating on the test fold.

\subsection{Model parameters} \label{sec:Model parameters}
The baseline DNN is a Multi-layer Perceptron (MLP) with three hidden layers of sizes: $\left[ 512, 1024, 512\right]$.
It has $1.1$M learnable parameters.
This model is fed with the $125$-dimensional (see Section~\ref{sec:Acoustic features}) mean feature vector of each audio segment.
The TDNN architecture adopted here is the same as~\cite{snyder2018x}, but we employ fewer kernels (or CNN filters) to reduce the model size.
Moreover, we use smaller \textit{statistics dimension} (see \textit{frame5} in the Table 1 of~\cite{snyder2018x}).
We experiment with two TDNN model sizes: 
\begin{enumerate}
    \item \textbf{TDNN-small}: $128$ filters at each CNN layer, $256$ as statistics dimension, total $\sim 280$k parameters.
    \item \textbf{TDNN-big}: $256$ filters at each CNN layer, $512$ as statistics dimension, total $\sim 954$k parameters.
\end{enumerate}
We incorporate batch normalization~\cite{ioffe2015batch} and $30\%$ dropout~\cite{srivastava2014dropout} for all the intermediate convolutional and linear layers of the DNN and TDNN models.
We also train a modified Resnet-18 model~\cite{he2016deep} to explore the learning capability of 2D time-frequency convolutions.
Two necessary changes are: usage of $32\times 4$ kernels for average pooling~\cite{he2016deep}, and having $4$ outputs nodes.
This model has $11.1$M trainable parameters.
We use Adam~\cite{kingma2014adam} optimizer for training with a batch size of $64$, learning rate of $0.001$, and $\beta_1=0.9, \beta_2=0.999$.

\begin{table}
	\centering
	\caption{Mean classification accuracy $(\%)$ of different models on segment-level acoustic scene classification task under $10$ fold cross validation. (FG = Foreground).}
    \begin{tabular}{cccc}
		\toprule
		\textbf{Model} & \textbf{\# parameters} & \textbf{FG active} & \textbf{Full} \\
		\midrule
		Baseline DNN & $1.1$M & 52.39 & 55.29 \\
		Resnet-18 & $11.1$M & 51.54 & 49.20 \\ 
		TDNN-small & $280$k & 56.80 & 56.22 \\
		TDNN-big & $954$k & 55.55 & \textbf{59.41}\\
		\bottomrule
	\end{tabular}
	\label{table:class_acc}
\end{table}
\section{Results and Discussion}
\begin{figure}
	\includegraphics[trim=0 235 0 240,clip,width=\linewidth]{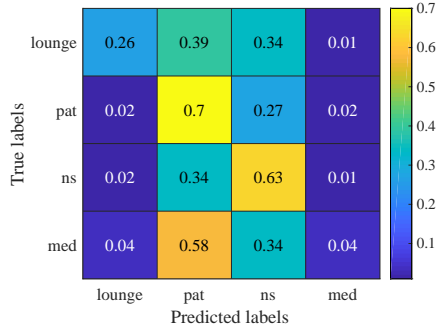}
	\caption{Mean test confusion matrix (values are in fraction of samples) for TDNN-big with no foreground speaker detection.}
	\label{fig:cmat}
\end{figure}
\subsection{Performance of the acoustic scene model}
Table~\ref{table:class_acc} shows the classification accuracy of all the models.
First, we analyze the performance of different models on the FG active data. 
We note that the baseline DNN performs better ($6.39\%$ absolute improvement) than the majority guess classifier (Section~\ref{sec:Acoustic scene label distribution}), which confirms the existence of acoustic scene patterns in the audio data.
Resnet-18 performs poorly for our task, possibly because of having too many trainable parameters with respect to the number of samples in the dataset (Section~\ref{sec:Effect of background audio}).
TDNN-small and TDNN-big outperform the baseline DNN by an absolute $4.41\%$ and $3.16\%$ respectively.

Next we move our attention to the full dataset.
Note that the number of samples is thrice of the FG active dataset.
Resnet-18 still shows poor generalization. 
TDNN-small gets a little improvement over that baseline (absolute $\sim 1\%$).
TDNN-big achieves an absolute boost of $4.12\%$ from the baseline DNN.

The results verify the efficacy of both the time delay networks to learn frame-level temporal dependencies even with fewer parameters compared to Resnet-18.
TDNN-small has much fewer parameters than the baseline DNN, yet it outperforms the baseline for masked and unmasked cases.
TDNN-big has similar number of parameters as the baseline, yet the former model shows better performance for both the data subsets.

The mean confusion matrix on test folds is shown in Figure~\ref{fig:cmat} for the best model \ie TDNN-big on the full dataset.
It is evident that the model is more accurate in predicting nurse stations (`ns') and patient rooms (`pat'), possibly because of having more training samples from these acoustic scenes.
The performance of the model degrades in predicting the lounge, with almost equal number of confusing samples coming from nurse stations and patient rooms.
The performance is poor for medication rooms (`med'), possibly because of having the least number of training samples (see Section~\ref{sec:Acoustic scene label distribution}).
Most of the confusions are originated from patient rooms, probably because of similar acoustic environment (comparatively quiet).



\begin{figure}
    \centering
    \begin{subfigure}[b]{0.49\linewidth}
        \includegraphics[trim=195 302 200 312,clip,width=\linewidth]{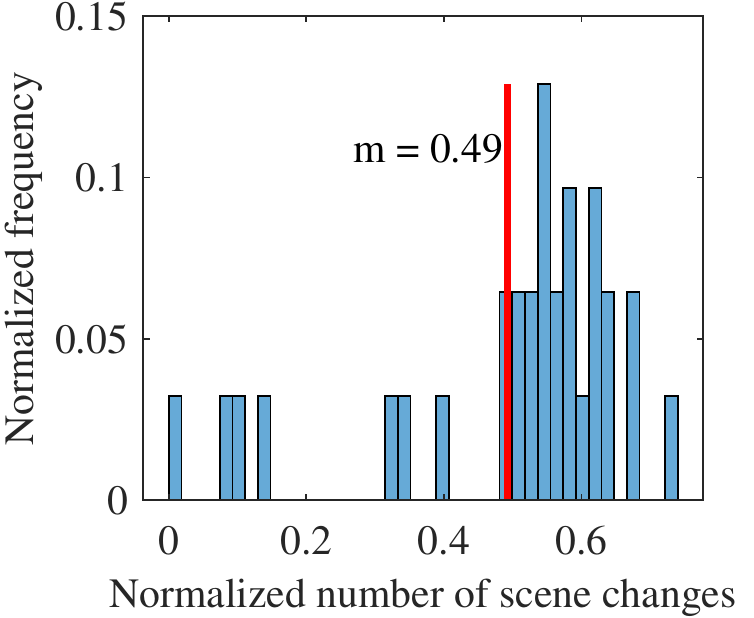}
        \caption{Day shift, true sequence}
        \label{fig:xcorrs}
    \end{subfigure}
    \begin{subfigure}[b]{0.49\linewidth}
        \includegraphics[trim=195 302 200 312,clip,width=\linewidth]{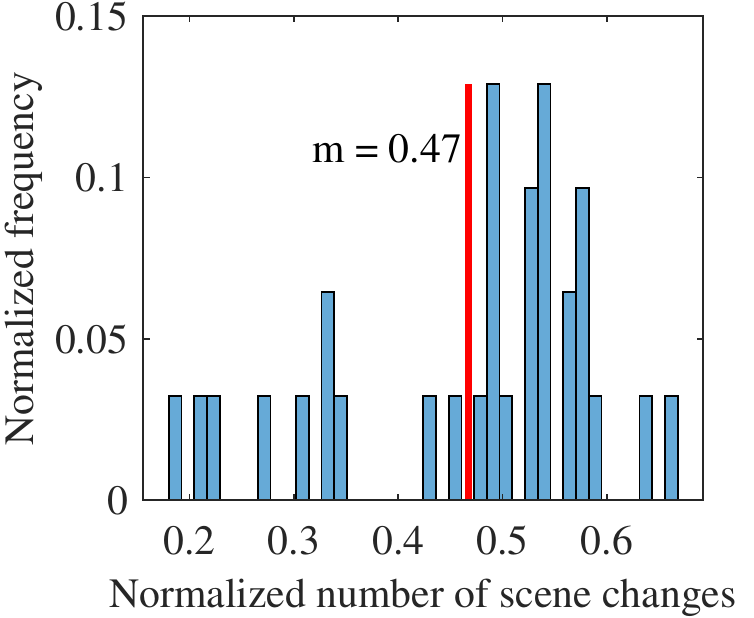}
        \caption{Day shift, predicted sequence}
        \label{fig:lags}
    \end{subfigure}
    ~\par\smallskip
    \begin{subfigure}[b]{0.49\linewidth}
        \includegraphics[trim=195 302 200 312,clip,width=\linewidth]{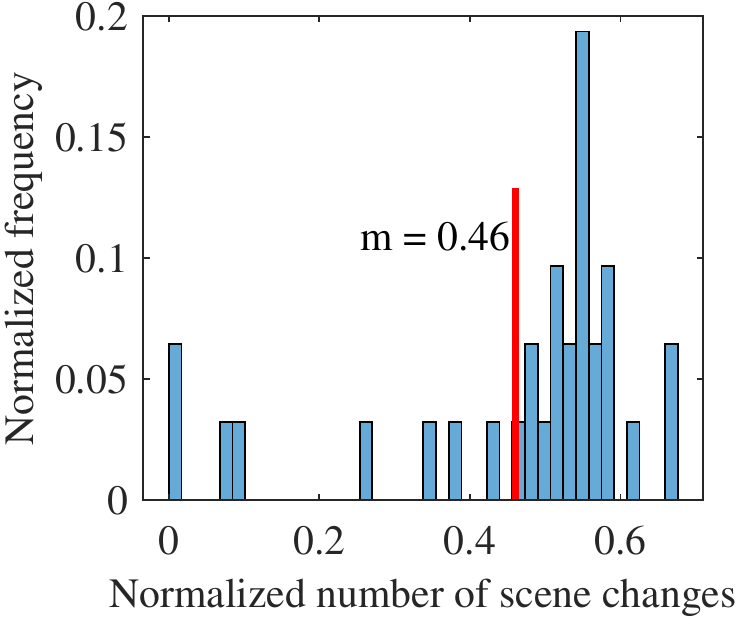}
        \caption{Night shift, true sequence}
        \label{fig:xcorrs}
    \end{subfigure}
    \captionsetup[subfigure]{font=footnotesize,labelfont=footnotesize}
    \begin{subfigure}[b]{0.49\linewidth}
        \includegraphics[trim=195 302 200 312,clip,width=\linewidth]{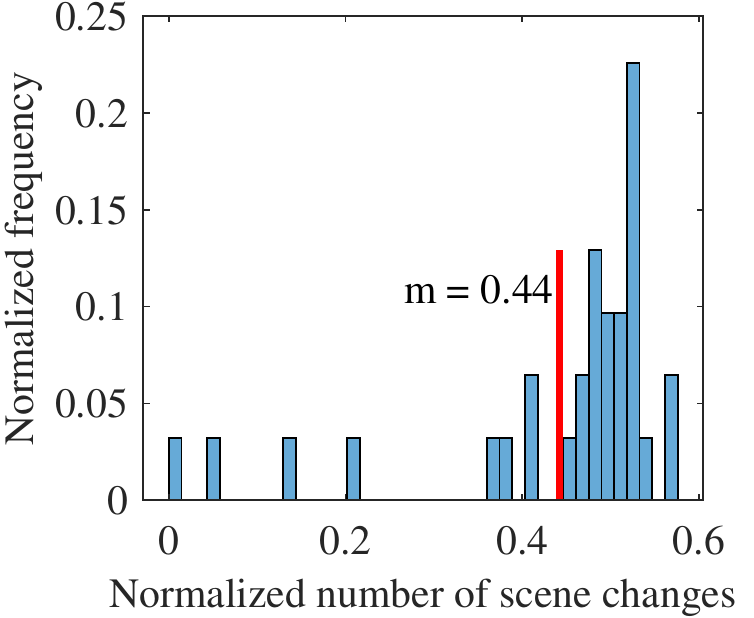}
        \caption{Night shift, predicted sequence}
        \label{fig:lags}
    \end{subfigure}
    \caption{Histograms for normalized number of changes in true and predicted acoustic scene sequences for day and night shifts.}
    \label{fig:loc_demographics}
\end{figure}

\subsection{Characterizing predicted sequence of acoustic scenes}
\label{sec:Characterizing predicted sequence of acoustic scenes}
Here we try to explore if patterns in changes in acoustic scenes experienced by a certain participant is related to the nature of  their job, specifically work shift and current position in the hospital.
We first try to find pattern in the true acoustic scene sequences, and then, verify whether the predicted scene sequences provide similar patterns.

Continuing the formulation of Section~\ref{sec:Characterizing dynamically varying acoustic scenes}, the histograms of normalized number of changes (see Equation~\ref{eq:nnz}) in the acoustic scene sequence (both for the true and the predicted sequences) are plotted in Figure~\ref{fig:loc_demographics} for day and night shift jobs.
The mean values are annotated on the histograms with red bars.
A quantitative analysis shows, for the true scene sequence, the mean of normalized number of changes ($1/{N_{\text{shift}}} \sum_{i=1}^{N_{\text{shift}}} \Delta \mathcal{Y}_i$ where $N_{\text{shift}}$ denotes number of participants in a particular shift) for day and night shifts are $0.49$ and $0.46$ respectively.
Similar, decrease in mean can be observed with the same metric for predicted sequence, $\Delta \mathcal{\hat{Y}}_i$ as well: 0.47 and  0.44 respectively for day and night shifts.

This distinction is more prevalent when we do similar analysis on current job positions in the hospital: nursing\footnote[3]{Includes registered nurse and nursing assitant.} \textit{vs.} non-nursing\footnote[4]{Includes monitor tech, physical therapist, occupational therapist, speech therapist, respiratory therapist, and other occupations.}.
For true acoustic scene sequences, the mean of normalized number of changes are $0.25$ and $0.54$ for non-nursing and nursing jobs respectively.
It also aligns with the intuition that the nursing jobs should be relatively more ambulatory.
More interestingly, the same metric for the predicted scene sequences are $0.34$ and $0.49$, \ie similar increasing trend for nursing jobs.

\section{Conclusions and future directions}
In this paper, we addressed the problem of predicting the acoustic scenes in a hospital workplace setting from long egocentric audio recordings.
The audio recordings obtained using wearable close-talk microphones capture both participant's speech and the ambient audio, and thus, it opens up new research directions in acoustic scene prediction through wearable or mobile devices.

We proposed a segment-level audio scene modeling to tackle extremely long audio recordings.
We employed a time delay neural network to model the acoustic features at the segment level.
The experiments showed that the employed time delay network performed the best in classifying the acoustic scenes, with or without an active foreground speaker detection step. 
Moreover, to characterize the egocentric view of acoustic scenes from the perspective of a participant, we explored the relationship between temporal pattern of the acoustic scenes experienced by the users with their job-types.

In the future, we will investigate how the egocentric acoustic patterns are related to individual mental states such as stress \cite{Hommedieu2019Lessons}.
The proposed segment-level modeling is additionally attractive, since it helps compressing the data, and thus higher layers of temporal systems can be designed for end-to-end learning.

\bibliographystyle{IEEEbib}
\bibliography{refs}

\end{document}